# PROBABILISTIC PATTERN RECOGNITION STUDY OF INTERACTION OF 1064 NM CW LASER WITH ICP PLASMA: ANTI-STOKES COOLING DUE TO ANALOGUE BLACK HOLE

M.F. Yilmaz[1], H.U. Rahman[2], F. Paletou[3], F Yilmaz[4], T. Malas[5], B. Karlik[6]

[1] Consultant to Magneto-Inertial Fusion Technology, Inc., Tustin, CA, USA

[2] Magneto-Inertial Fusion Technology, Inc., Tustin, CA, USA

[3] Université de Toulouse, Observatoire Midi–Pyrénees, CNRS, CNES, IRAP, F–31400 Toulouse, FRANCE

[4] Department of Computer Sciences, New Jersey Institute of Technology, New Jersey, NY, USA

[5] Lorentz Solution Inc., Fremont, CA, USA

[6] Neurosurgical Simulation Research and Training Centre, Department of Neurosurgery, Montreal Neurological Institute and Hospital, McGill University, Quebec, CANADA

**ABSTRACT**

Pattern recognition and machine learning techniques are known to play an emerging role for studying underlying physics of light-matter interaction. While on the subject, photon condensation and Anti-Stokes cooling remain compelling phenomena of laser- underdense plasma interaction.In this work, the interaction of a 1064 nm continuum-wave laser with inductively-coupled plasma (ICP) of Mercury(Hg) has been studied by probabilistic pattern recognition over observed time resolved spectral data.3D vector fields obtained by the probabilistic pattern recognition over spectral database proficiently illuminates the analogue black hole (ABH) and its sonic horizon in which homoclinic orbit Whistler waves are exited. Vector spectra show the collective behaivor of condensation and anti-Stokes cooling is due to the electronic transitions of Hg1 trapped by the phonon sink of ABH of laser. Modeling of subsonic phonon vector spectrum obtained by the probabilistic linear discriminant analysis(PLDA)  estimates the 0.70 nanoKelvin of temperature of the sink region of ABH.

**INTRODUCTION**

Plasmas are tunable mediums, which, in turn, can act as absorbers, transmitters or reflectors, depending on the frequency range and application of interest. Therefore, plasmas are known to have a widespread use in industry for several applications, such as stealth technologies in radar applications and radio communications. In this context, the propagation of radio-frequency (RF) electromagnetic waves in uniform, non-uniform, magnetized or unmagnetized plasmas have been studied extensively both experimentally and theoretically [1]–[4]. It has been shown that it is possible to obtain high levels of absorption in a broadband range including RF and microwave frequencies [5]–[7] by adjusting some plasma parameters, such as magnetic field strength and plasma density.

[1] Independent Scholar, Fremont, CA, USA. *email: fthyilmaz53@gmail.com

However, the propagation of low-intensity laser light within cold plasmas have found less interest in comparison to RF or microwave electromagnetic waves emitted from antennas [4], [8]–[10]. Although laser is indeed a special form of electromagnetic wave with extremely high frequency and propagation of laser can be best described by the wave model, a plausible explanation of the absorption mechanism needs the particle model. For instance, the Drude model elucidating the interaction of RF waves within plasma has failed to explain the loss mechanism of plasma after the introduction of a laser beam [8]. Laser cooling of nanoparticles and semiconductors has been studied in detail and most of the studies suggest the cooling is attributed to photon upconversion and anti-Stokes shifting, especially for the room temperature mediums [10 and 11]. It has been shown that anti-Stokes shifts are due to thermal (phonon) interaction with the excited atoms [12]. Betz et al., 2012 showed that hot electrons were cooled by the inverse Lorentz profile (sinking mode) of acoustic phonons [13]. Besides, Mario et al., 2002 stated that phonon sinks have been observed in analog or so-called acoustic black holes (ABH) and these condensate phonon flows of ABH continually inward and sinks at r=0. It is well known that such sinks can also be generated by means of an out-coupler laser beam at the origin [14].

In this study, we investigate the cooling of Hg plasma by means of anti-Stokes polarization of photons in which electronic transitions of Hg (@ 437 nm and 546 nm) trapped through phonon sink of ABH that is revealed by probabilistic pattern recognition techniques over time-resolved spectra of cold Hg plasma. Principal component analysis (PCA) and linear discriminant analysis (LDA) are well-known pattern recognition methods for processing and carrying out dimension reduction of the data. PCA and LDA have found application in several areas, such as medicine, robotics and remote sensing. They have also been found in many applications in spectroscopy, especially in unmixing species and decomposing overlapped spectral lines of UV-VIS-NIR spectroscopy to extract the spectral fingerprints. In addition, they are used in the spectroscopy of astrophysical plasmas and laboratory plasmas for extracting the plasma parameters and compositions of ion species [15]–[21]. While PCA does identify the linear subspace in which most of the data's energy is concentrated, LDA identifies the subspace wherein data between different classes is most spread out. This, in turn, makes LDA suitable for recognition problems such as classification. In the meantime, PCA and LDA have limitations related to dimensionality reduction, such as nonparametric representation of data and high computational cost. Tipping et al. 1999 and Ioffe et al. 2006 posited that probabilistic model of PCA can make nonlinear dimension reduction without computing the covariant matrix in order to manage computation time. Furthermore, probabilistic PCA can successively represent the missing or unseen data, since the model based on transformation low dimensional subspace to higher dimensional subspace using Gaussian-latent variable model [22 and 23].

In this study, the interaction of continuum mode 1064 nm diode laser light with the ICP Hg plasma is analyzed both experimentally and theoretically. The experimental part is premised on the power-meter measurements and spectroscopy. PPCA and PLDA over UV-Vis spectra of Hg plasma is used as a feature extractor to investigate the polarization, scattering structures and electron temperature, as well as density of the plasma in the presence of different powers of laser beam. Similarly, the theoretical component is premised on the study of the PPCA and PLDA analysis and non-local thermal equilibrium (non-LTE) radiative modeling of experimental spectra both in the absence and presence of laser beam.

## METHODS

### Experiment

The uniform magnetized plasma slab has been generated by inductively coupling a 13.56 MHz RF generator (60 Watts) to the fluorescent light bulb, which has a length of 30 cm and a diameter of 2.0 cm. The spectra have been recorded by a charge-coupled (CCD) spectrometer device of AvaSpec-ULS3648. Continuum-wave (CW) diode laser lights at different powers was directed to the plasma, after which power levels were recorded by means of a Thorlab-PM100D power-meter. Fig. 1a illustrates the laser powers and corresponding absorbance coefficients that are obtained using standard loss medium model. It can be seen that low power laser attenuated quickly, and that the plasma becomes transparent to the laser beam as the power of laser increases. Fig. 1b shows the experimental spectra of Hg plasma in the absence and presence of laser light at different powers.

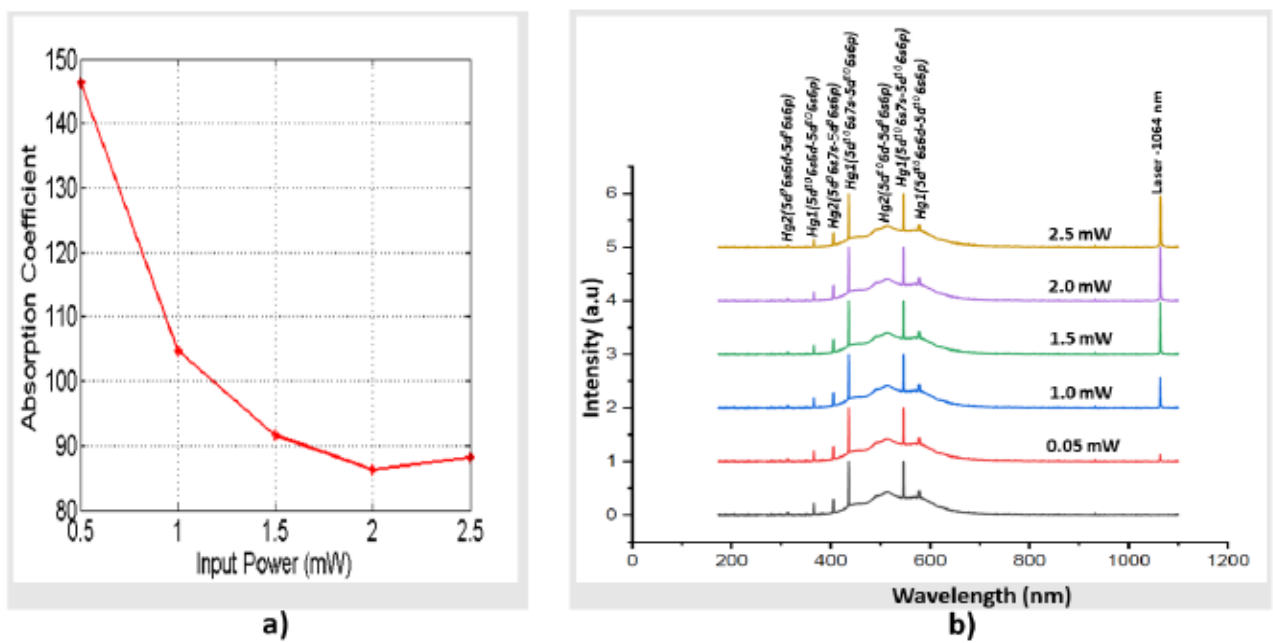

Fig. 1 a) Absorption coefficients at different laser powers and b) UV and UV-Vis spectra of Hg plasma in the absence and the presence of the laser light.

**Probabilistic principal component and linear discriminant analysis**

The Principal Components Analysis (PCA) is a linear transformation employed to extract the subspace of a dataset having the maximum variance. Due to the nature of the model selection criteria and information complexity criterion, the essential component of applying an information criterion for a model is a maximum likelihood estimator. However, classical principal component analysis lacks a probability model that is necessary to derive maximum likelihood estimation. Tipping and Bishop et al., 1999 have proposed a latent variable model that is closely related to factor analysis. In this model, named the Probabilistic Principal Component Analysis (PPCA), the maximum likelihood estimation of the parameters can be used to construct the fundamental axes of a set of observed data vectors. As the name of this model implies, PPCA is the probabilistic version of the well-known PCA method. PPCA basically uses a Gaussian-latent variable model that is related to Bayesian statistical factor analysis. Let $y_1, y_2,...,y_k$ be observable variables with the means $u_1, u_2,..,u_k$. We assume that each random variable can be written as a linear combination of the latent variables $x_1, x_2, ..., x_l$ where $l < k$ (common factors) plus error term, i.e.

$$y_i - u_i = w_{i1}x_1 + w_{i2}x_2 + \cdots + w_{il}x_l + \varepsilon_i \quad (1)$$

where $w_{ij}$s are loadings. Assuming to have $n$ observation, we can write the following matrix equation:

$$Y = WX + U + \varepsilon \quad (2)$$

Where $W$ is $kxn$ matrix of loadings. So, the matrix of loadings relates observed variables to latent variables, and $U$ allows the model to have nonzero mean.

By Gaussian noise model $\varepsilon \sim N(0, \sigma^2 I)$, where $I$ is the identity matrix, the conditional probability distribution of $Y$ given $X$ turns out to be

$$P(Y|X) \sim N(WX + U, \sigma^2 I) \quad (3)$$

Assuming normal distribution of latent variables $X \sim N(0, I)$ gives observable variables are also normally distributed and

$$Y \sim N(U, WW^T + \sigma^2 I) \quad (4)$$

In this model, Principal components are given by maximum likelihood of estimates of parameters which are $W$ are $\sigma^2$. Note that PCA is a specific case of probabilistic PCA as $\sigma^2$ tends to 0. Eigen decomposition of the sample covariance matrix or EM(Expected maximization) algorithm can be employed to calculate Maximim likelihood estimates of parameters. For further detail, one can read Tipping et al., 1999 [21],

Linear Discriminant Analysis (LDA) distinguishes the data in the data set in accordance with the properties of the data when it is assigned to variable groups. The discriminant functions obtained through discriminant analysis consist of linear components of estimation variables. LDA confronts challenges when the discriminative information is not in the means of classes and small sample size problem. PLDA model is the probabilistic version of LDA wherein both within-class and between-class variances are represented as multidimensional Gaussians. It is possible to build a model of a previously unseen class using a single example with PLDA. Moreover, it is easy to combine multiple examples for a better representation of the class. Let us assume $x_1, x_2,...,x_k$ be feature vectors for $n$ classes. We assume each feature vector can be decomposed as

$$X = U + Lw + Fz + \epsilon, \qquad (5)$$

Where $U$ is the average feature vector, L is the matrix whose columns form the basis for the identity space, w is identity latent variable, F is the matrix whose columns form a basis for within-class subspace and $\epsilon$ is the noise term. Latent variable vectors are assumed to be independent and normally distributed. Noise term is also assumed to be gaussian with mean zero and diagonal covariance matrix $\Sigma$. Although the matrices L and F are very close to within class and between class matrices in classical LDA, they are calculated probabilistically in this model. The likelihood that two feature vector belongs to the same class is modeled by the following equation:

$$P(Y|w, z, U, L, F, \Sigma) \sim N(U + Lw + Fz, \Sigma) \qquad (6)$$

In this probabilistic model, The aim is to find the maximum likelihood estimates of the parameters U, L, F, $\Sigma$ by training a large dataset. Maximum likelihood estimation is a tool that uses the likelihood function to estimate the parameters of a probability distribution for which the given dataset is most possible for the given statistical method. Since the latent variables are not known the maximum likelihood estimates of these parameters are performed iteratively by using expectation maximization (EM) algorithm. A more detailed explanation can be found in studies conducted by Ioffe et al., 2006 and Prince et al., 2007 [22 and 23].

In the present work, PCA is applied to the set of 60 spectra of size 2831x1 for the laser powers of 0.0, 0.05, 1.0, 1.5, 2.0 and 2.5 mW. Principal components are eigenvectors of the covariance matrix consisting of the covariances between different variables and hence, is a matrix of size 2831x2831. Therefore, each vector (principal component) also has a size of 2831x1. Then original data is projected onto the space spanned by the first three, most dominant, principal components denoted by |PC1>, |PC2> and |PC3>. As the first step, *PCA* is used to reduce the size of within-class scatter matrix ($S_w$) for each electron beam fraction so that it can be inverted. The data set comprises of 6 laser powers (0.0, 0.05, 1.0, 1.5, 2.0 and 2.5 mW) × 60 spectra of size 2831 × 1. By applying *PPCA*, the dimension is reduced to 30 by projecting each of the spectra into the space generated by the most dominant 30 PCs. As the second step, *PLDA* is applied to the 6 classes of powers where each class consists of 360 vector of size 2831 × 1, with the dimension being reduced to 3. The new vector space is generated by |PLD1>, |PLD2> and |PLD3>.

**RESULTS**

3D plots of PC1, PC2 and PC3 coefficients are illustrated in Fig. 2, which demonstrates that the increase of laser power causes the plasma to be more collective and linearizes except for the 0.5 mW of laser power. |PC1> vector spectra in Fig. 3a correspond to the polarization due to E-field of laser, whereas Fig. 3b is the zoom in of |PC1> vector spectra. This shows the enhancement of intensities (Stokes-I

polarization profiles) by increasing the laser power. |PC2> vector spectra in Fig. 4a correspond the polarization due to B-field of the laser and Hg1 *(5d$^{10}$6s7s-5d$^{10}$6s6p)* line follows the circular polarization profiles of Stokes-V. Furthermore, Fig.4b illustrates the zoom in of spectral line of Hg2, while the peak of the line of Hg2 experiences the Anti-Stokes shifting by increasing the laser power. Intensity of the line profiles moves to the left and decreases (Anti-Stokes) with an increase in the laser power (Fig. 4b) [24 and 25].

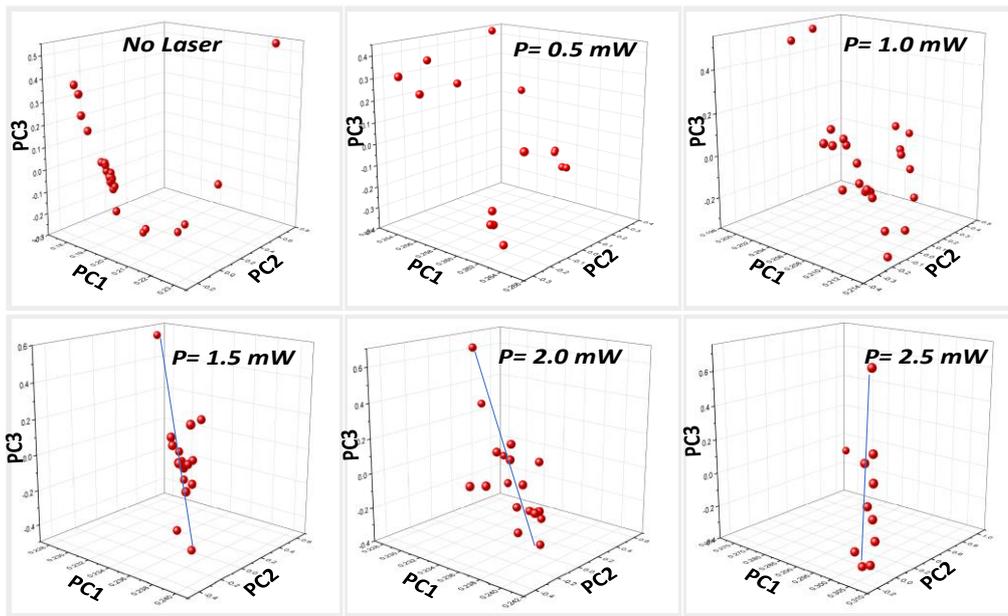

Fig. 2 3D representations of PC of coefficients with increasing laser power.

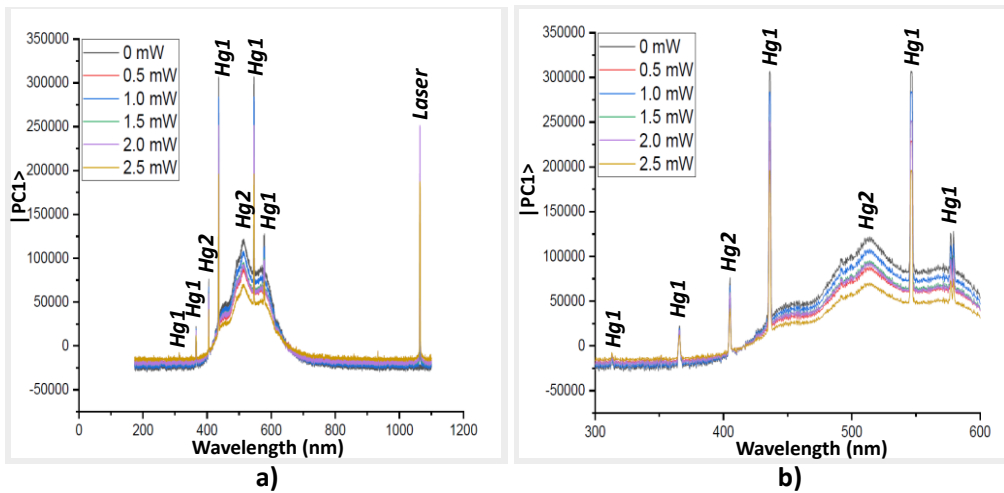

Fig. 3 a) |PC1> vector spectra of Hg with increasing laser power. b) Stokes profiles (zoom in of |PC1> spectra).

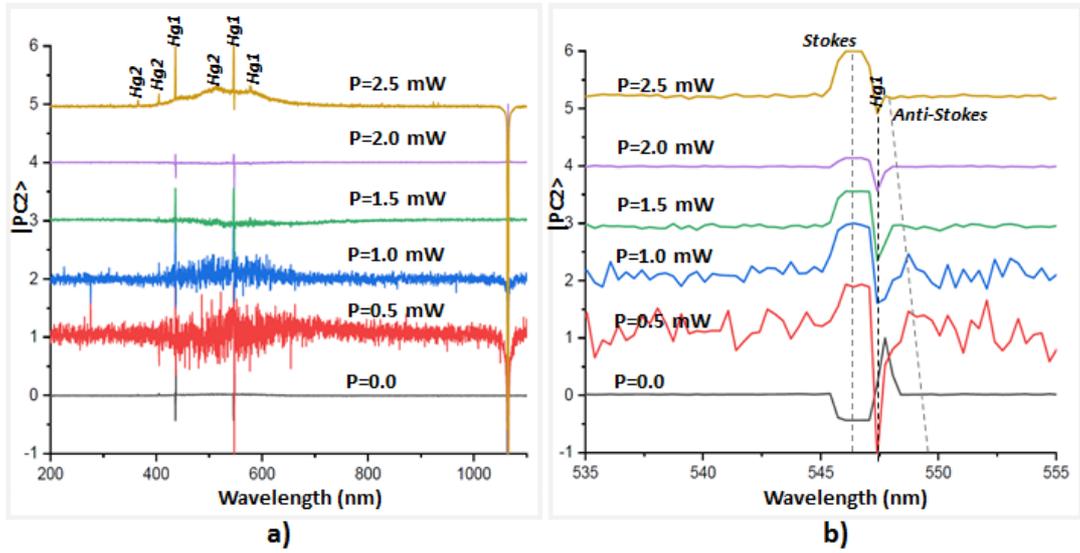

Fig. 4 a) |PC2> vector spectra of Hg plasma in the absence and presence of laser beam with different powers. b) Anti-Stokes, resonant and Stokes of Hg1 profiles (zoom in of |PC2> spectra).

Plasma electron temperature modeling of experimental spectra is realized by SPARTAN radiative transfer code. SPARTAN uses the non-thermal plasma approach that follows the thermal equilibrium and velocity distributions of ions as well as non-Maxwellian distributions of electrons [26]. SPARTAN generated synthetic spectra in Fig. 5a demonstrates that the intensity of all considered transitions increases with an increase in the plasma electron temperature. Modeling of Hg plasma with synthetic model in the absence of laser suggests the plasma electron temperature of 0.85 eV. Besides, using the ratio of anti-Stokes to Stokes intensity is another method to observe the trends of plasma electron temperature in experimental spectra [25]. For that reason, we have used the Anti-Stokes—Stokes intensity ratios of transition of $Hg1(5d^{10}6s7s-5d^{10}6s6p)$ at 546 nm to obtain the trends of electron temperatures with the corresponding laser powers (Fig.4b). Fig. 5b shows that plasma is only heated to 7 eV by the 0.5 mW of laser power than cooled by the increasing the laser power.

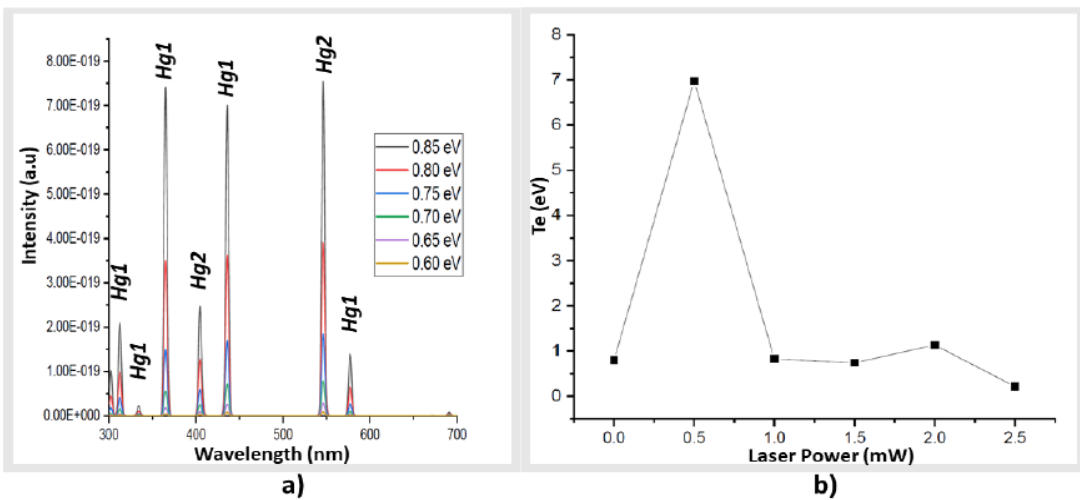

Fig. 5. a) Plasma electron temperature dependance of non-LTE synthetic spectra at electron density of 1x $0^{13}$cm$^{-3}$ b) Modeling of plasma electron temperature with corresponding laser power.

|PLD1>, |PLD2> and |PLD3> vector field spectra are illustrated in Fig. 6.a. |PLD1>, |PLD2> and |PLD3> correspondingly denote down scattered, up scattered photon and phonon spectrum. Down shifted spectrum has higher intensities and electron-phonon spectrum has relatively lower intensities [27]. Electron-phonon hybrid spectrum (|PLD3>) has been illustrated in Fig.6.b. with the presence of the supersonic and subsonic(laser) regions and absorptions of Hg1 electonic transitions at 437 nm and 546 nm and. Zoom in spectrum of |PLD2> and |PLD3> for the Hg1 transition clearly shows the Anti-Stokes and Stokes shifts. Electron oscillations spectrum (|LD4>) are represented in Fig. 6.d. Modeling of electron oscillations using fast Fourier modeling suggests that the electron oscillations is around 166 Hz (0.16 kHz). The plasma electron density ($n_e \sim 4 \times 10^{13}$ cm$^{-3}$) is obtained using the Eq.7, where *$w_{pe}$* denotes the electron frequency and *$e_0$* signifies the permittivity of the free space [30, 31 and 32].

$$\omega_{pe} = \sqrt{n_e e^2 / \varepsilon_0 m_e} \qquad (7)$$

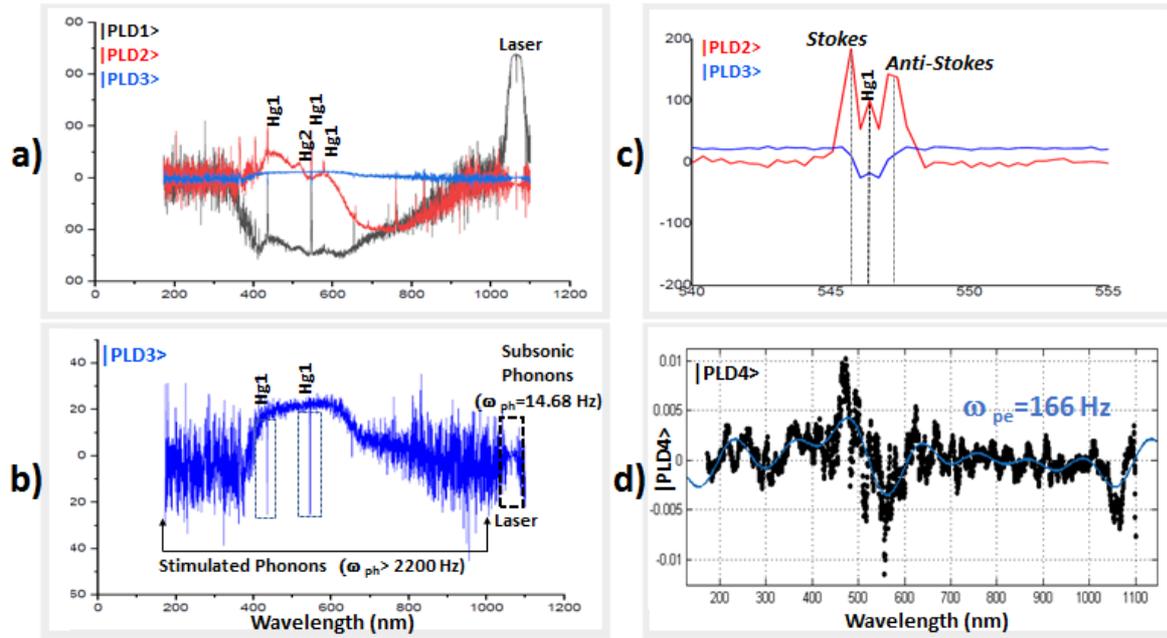

Fig. 6 a) Up scattered(|PLD1>), down scattered(|PLD2>) photon, electron-phonon(|PLD3>) vector spectra of Hg plasma b) electron-phonon spectrum(|PLD3>) c) Zoom in spectra of Anti-Stokes,resonant and Stokes of Hg1 from |PLD2> and |PLD3> d) electron (|PLD4>) spectrum and its FFT modeling

The surface-plot of 3D representation of PLDA vector field spectra has been illustrated in Fig.7a which exhibits the acoustic (analog) black hole with phonon sink, sonic horizon and whistler wave formation [28 and 29]. While on the subject, Fig.7b shows the typical convex ABH diagram which has supersonic and sonic flow regions with the sonic horizon [33] . Shannon entropy of |PLD1> ,|PLD2> and |PLD3> spectra in 3D plot that sinking region has higher entropy when compared to plasma region (Fig. 7c) [34]. Zoom in of phonon (|PLD3>) spectrum between 1050-1080 nm corresponds sinking region (laser light-1064 nm) of ABH. It shows that one phonon absorbed and the other is emitted in sinking region. Fast fourier modeling gives the phonon frequecy of 14.68 Hz which estimates the 0.70 nanoKelvin of temperature (Fig. 7d). Furthermore, one phonon is absorbed (pulled) and another one is emitted (pushed) at 1064 nm [35]

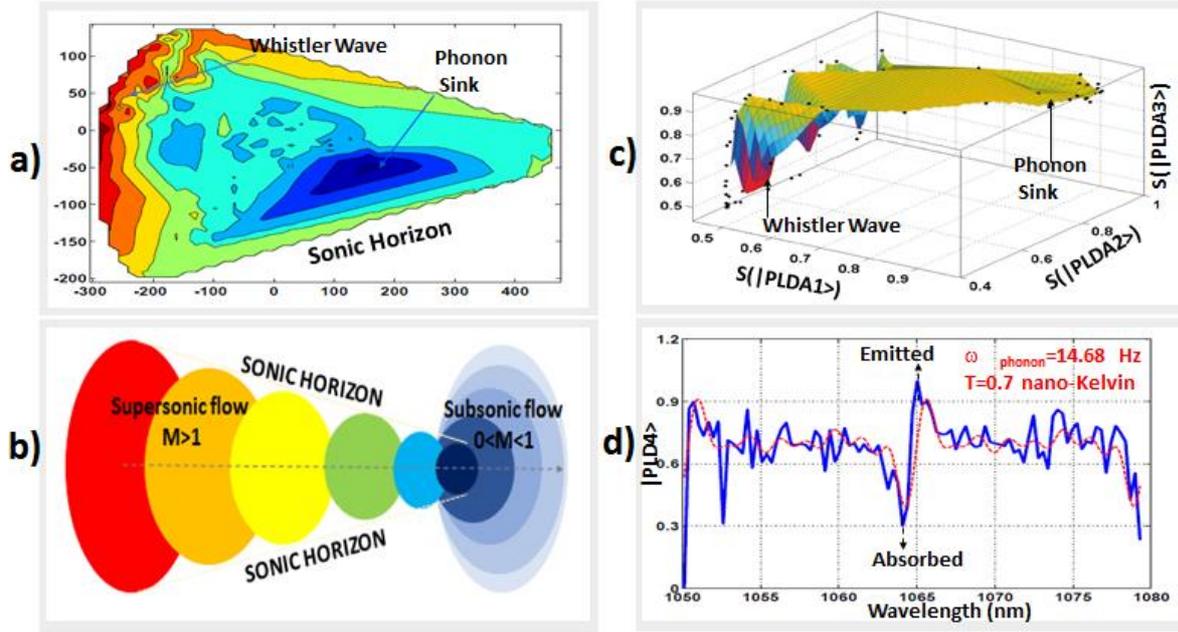

Fig. 7 a) Acoustic black hole- topological plot of 3D representation of PLD vectors. b) A schematic presentation of acoustic black hole with convex geometry. c)A schemactic presentation of 3D Shannon entropy of |PLD> vector spectra. d) Fast fourier modeling of phonon spectrum( |PLD3>) in the laser (phonon-sink) region-1064 nm.

**DISCUSSION**

Laser light can cool, trap and condensate the atoms/molecules. Leduc et al.2009 stated that when particles are condensate by the laser or magnetic cooling, interparticle distance become smaller and particles follows the collective behavior [36]. Plasma condensation is usually pronounced in coronal plasmas and realized when magnetic force dominates the gravitational and the pressure forces [37]. O'Shea et al. 2007 has showed that the enhancement of the intensity of radiation and strong electric fields in coronal loops, has been depicted by the stability of the plasma condensation [38]. Collective behavior of plasma associated with the cooling of electron temperatures in our data agrees with the fact that plasma is condensate by the increase of laser power with the exception of 0.5 mW.

Light propagates in plasma medium following the dispersion relation $\omega^2=\omega_{pe}^2/<\gamma>+k^2c^2$, where $\omega$ signifies the frequency of light, $\omega_{pe} = \sqrt{n\varepsilon_e e^2/\varepsilon_0 m_e}$ shows the plasma frequency and the average $<\gamma>$-factor. For lasers $\omega \gg \omega_{pe}$ holds in general, and the wave number becomes almost identical to that of the free space [32]. This amounts to no reflection at the boundaries and scattering, thus causing loss of energy. Stimulated Brillouin scattering of intense laser radiation in plasma is one of the most important parametric process, which describes the decay of the incident laser radiation into the scattered electromagnetic wave as well as an ion acoustic wave-phonon (Eq.8) [39].

$$\text{Photon} \rightarrow \text{Whistler (ion acoustic-phonon)} + \text{Scattered photon} \tag{8}$$

Since the inter-class and in-class variances (scattering matrixes) are represented as multidimensional Gaussians in PLDA, one can identify bound states, virtual states or resonances of the energy poles of the scattering matrix. Fig.6.c shows that down and up scattered spectra and phonon spectrum in Fig. 6.a are realized in the sonic zones of ABH to gradually to form of Whistler waves. It is well known that ABH is made of acoustic waves and it is formed inside the Bose-Einstein Condensate (BEC) state [28 and 29]. ABH concept was developed by Unruh et al., 1981 from the analogy between general relativity and super fluid dynamics. In an ABH, the velocity of light is substituted by the velocity of sound, and event horizon is subsituted by sonic horizon which separates the supersonic and subsonic regions. (Fig. 7b). Besides , in an ABH sound waves are swept along by supersonic flow in which fluid velocity exceeds the velocity of sound [33]. Steinhauer et al. 2014 constructed ABH in BEC state and observed that very weak quantum fluctuations occur to produce virtual phonon pairs even in frigid BEC state [29 and 40]. Such fluctuations create the gradient of air pressure. Steinhauer et al.,2014 also stated that virtual phonons pop up into existence across the condensate's event horizon. At the end, one is absorbed by the supersonic sink and the other escaping away from it, which exhibits the same analogy of Hawking Radiation [35 and 41]. Our results in Figure .7d shows that one phonon is absorbed and the other one is emitted in the sinking region. The temperature (~ 0.70 nanoKelvin) is estimated from Fourier modeling of the phonon frequency spectrum (14.68 Hz) which also agrees with the Steinhauer's observations.

Whistler waves have been observed in under dense plasmas with very low frequency propagation along a magnetic field B. It is circularly polarized and occurs when plasma frequency is less than that of the electron cyclotron frequency. The interaction between whistler waves and lower hybrid waves (electron-phonon) in a plasma are still among the compelling phenomena.Whistler waves are capable of destabilizing a magnetized plasma by exciting the lower hybrid wave together with the ion acoustic wave (phonon). This excitation is parametric and follows cooling by the energy loss transferred from resonant to non-resonant wave. Finally, non-resonant waves conclude to chaotic wave amplitudes which can stimulate the phonons to supersonic frequencies (Fig.6b) [42, 43 and 44]. Fig.6c and 7a indicates the trapped phonons flows towards to center of ABH. ABH sweeps away the resonant transitions of Hg @ 436 nm and 546 nm, which also maintain the cooling and condensation of the plasma. Furthermore, Rabinovich describes the Whistler waves as homoclinic chaotic strange attractor generated by bifurcation in which a periodic orbit collides with the saddle equilibrium point (resonant) to itself [45 and 46]. On the other hand, B.Karlik and M.F.Yilmaz et al., 2020 have recently shown that 3D vector fields obtained out of spectral database of UV light-nanoparticle interaction using linear pattern recognition can successively reveal the strange attractors[47].

The attenuation of laser intensity $I$ can be modeled by $\frac{dI}{dx} = -\kappa_{ib} I$, where $\kappa_{ib}$ is the inverse bremsstrahlung absorption coefficient. If we denote $\rho_c$ as the mass density of the plasma corresponding to the critical density, absorption coefficient is shown to scale as follows:

$$\kappa_{ib} \propto \left(\frac{\rho}{\rho_c}\right)^2 Z\lambda^{-2}T_e^{-3/2}\left(1-\frac{\rho}{\rho_c}\right)^{-1/2} \quad (9)$$

Fig. 8.a shows the absorption coefficient versus $Te^{-3/2}$ which follows the almost inverse trend shown in Fig.1. The absorption coefficients obtained in Fig.1 by power measurements reveal that there is a significant absorption in the plasma with the laser power of 0.5 mW and decrease exponentially by increased laser power [48]. Fig.8b shows the resonance of E and B field of the laser light at 0.5 mW obtained by PC vector spectra. This confirms the fact that the plasma's heating is provided by the resonant absorption. Resonance absorption observed when a p-polarized light shine on a plasma with density gradient (ω=ω$_{pe}$) in which polarized light resonantly excites an electron plasma wave at critical density. Further increasing laser power results cooling of the plasma.

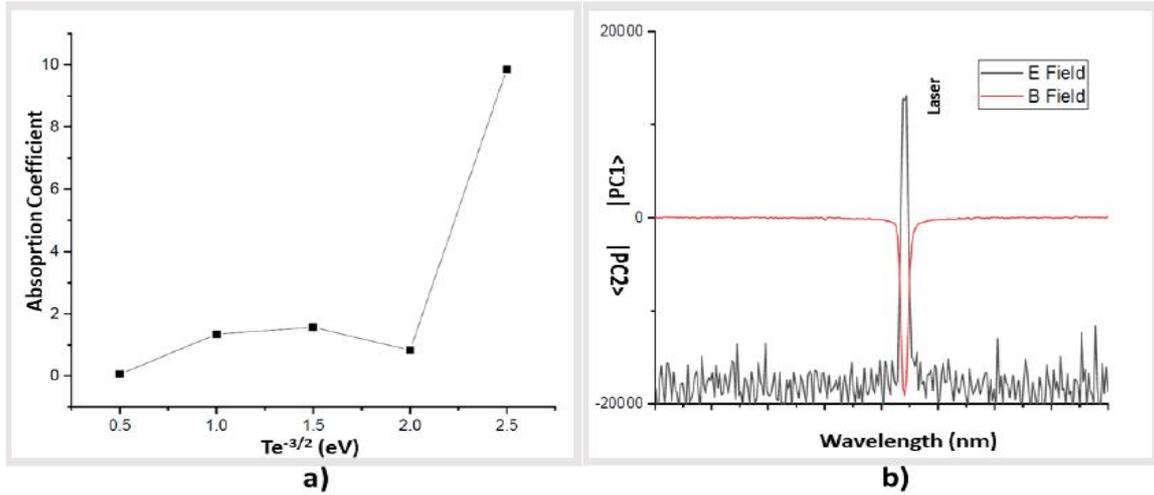

Fig. 8. a)Absorption scaling of laser plasma interaction at different power levels. b) Resonance spectrum ($\overrightarrow{|E|} \cong \overrightarrow{|B|}$) of laser spectra at 0.5 mW obtained by PC vectors

**CONCLUSION**

In this study, CW laser propagation in low temperature magnetized collisionless Hg plasma has been investigated by means of probabilistic pattern recognition over emission spectroscopy. Since PPCA and PLDA extends PC and LD vectors into higher dimensions, PPCA and PLDA of spectra can be powerfull tool to reveal more underlying physics of the laser plasma interaction. PPCA of spectral data can distinct the Stokes and anti-Stokes polarizations due to E and B fields correspondingly and exhibits the collective behaivour of plasma. 3D PLDA vector spectra succesfully identifies supersonic and subsonic regions of ABH and segregate the upshifted and down shifted spectra and hybrid electron-phonon spectrum, correspondingly. The spectroscopic modeling of plasma suggests that electron temperature is 0.85 eV and electron density be $4\times10^{13}$ cm$^{-3}$ in the absence of laser light, and the temperature is increased to ~ 7 eV at only 0.5 mW of laser ower due to resonant absorption. However, increased laser power cools the plasma, but more collective is indicative of plasma condensation. Pattern recognition over the spectra reveals that phonon sink assisted Anti-stokes shifting is the main mechanism of the cooling of plasma.

**ACKNOWLEDGEMENTS**

This research was funded by The Scientific and Technological Research Council of Turkey (Tubitak-EEAG-113E097).

**LIST OF REFERENCES**

1.  Jin F., Tong H., Shi Z., Tang D., and Chu P. K, Effects of external magnetic field on propagation of electromagnetic wave in uniform magnetized plasma slabs. Computer physics communications, 175, 8, pp.545-552. (2006).

2.  Tang D. L., Sun A. P., Qiu X. M., and Chu P. K., "Interaction of electromagnetic waves with a magnetized nonuniform plasma slab," IEEE Transactions Plasma Science, 313, pp. 405–410. (2003)

3.  Sheffield J., Froula D., Glenzer S. H., and Luhmann Jr N. C., Plasma scattering of electromagnetic radiation: theory and measurement techniques. Academic Press (2010).

4.  Kruer W. L., The physics of laser plasma interactions. Boulder, Colorado, Westview. (2003)

5.  Vidmar R. J., "On the use of atmospheric pressure plasmas as electromagnetic reflectors and absorbers', IEEE Transactions Plasma Science, 18, 4, pp.733–741. (1990)

6.  Aydin K., Ferry, V. E., Briggs R. M., and Atwater H. A., Broadband polarization-independent resonant light absorption using ultrathin plasmonic super absorbers, Nature Communication., 2, pp.517. (2011).

7.  Yuan C.X., Zhou Z.X., Zhang J. W., Xiang X.L., Feng Y., and Sun H.G., Properties of propagation of electromagnetic wave in a multilayer radar-absorbing structure with plasma-and radar-absorbing material, IEEE Transactions Plasma Science, 39,9, pp.1768–1775. (2011)

8.  Mora P., Theoretical model of absorption of laser light by a plasma, Phys. Fluids, 25, 6, 1051. (1982)

9.  Wang Y. and Zhou Z., "Propagation characters of Gaussian laser beams in collisionless plasma: Effect of plasma temperature," Physics of Plasmas, 18, 4, pp.43101. (2011)

10. Muys P., Stimulated radiative laser cooling. Laser Physics, 18, 4, pp.430-433. (2008)

11. Djeu N. and Whitney W. T, Laser cooling by spontaneous anti-Stokes scattering. Physical Review Letters, 46, 4, 236. (1981)

12. Nemova G., & Kashyap R. Laser cooling of solids. Reports on Progress in Physics, 73, 8, pp.086501. (2010).

13. Betz A. C., Vialla F., Brunel D., Voisin C., Picher M., Cavanna A., ... & Pallecchi E. Hot electron cooling by acoustic phonons in graphene. Physical Review Letters, 109,5, pp.056805. (2012)

14. Mário, N., Visser M., and Volovik G. E. Artificial black holes. World Scientific, 2002.

15. Prince S. J., and Elder J. H., Probabilistic linear discriminant analysis for inferences about identity. In 2007 IEEE 11th International Conference on Computer Vision, pp. 1-8. (2007)

16. Jolliffe I. T., Principal component analysis. (2002)

17. Balakrishnama S. and Ganapathiraju A., Linear discriminant analysis-a brief tutorial. Institute for Signal and Information Processing, 18, pp.1-8. (1998)

18. Luthria D. L., Mukhopadhyay S., Robbins R. J., Finley J. W., Banuelos G. S., and Harnly J. M., "UV Spectral Fingerprinting and Analysis of Variance-Principal Component Analysis: a Useful Tool for Characterizing Sources of Variance in Plant Materials," J. Agric. Food Chem, 56, 14, pp. 5457–5462. (2008).

19. Paletou F., A critical evaluation of the principal component analysis detection of polarized signatures using real stellar data. Astronomy & Astrophysics, 544, A4. (2012)

20. Larour J., et al, "Modeling of the L-shell copper X-pinch plasma produced by the compact generator of Ecole Polytechnique using pattern recognition." Physics of Plasmas 23.3, 033115. (2016)

21. Tipping M. E. and Bishop C. M, Probabilistic principal component analysis. Journal of the Royal Statistical Society: Series B (Statistical Methodology), 61, 3, pp.611-622. (1999)

22. Ioffe S., Probabilistic linear discriminant analysis. In European Conference on Computer Vision Springer, Berlin, Heidelberg, pp.531-542. (2006)

23. Simon Prince J.D. and Elder J. H, "Probabilistic linear discriminant analysis for inferences about identity." IEEE 11th International Conference on Computer Vision. (2007)

24. Wiehr E., Spatial and temporal variation of circular Zeeman profiles in isolated solar Ca (+) K structures. Astronomy and Astrophysics, 149, pp.217-220. (1985)

25. Goldstein T., Chen S., Tong Y., Xiao J., Ramasubramaniam A., and Yan, J., Raman scattering and anomalous Stokes–anti-Stokes ratio in MoTe 2 atomic layers. Scientific reports, 6, pp.28024. (2016)

26. Da Silva M. L, Vacher D., André M. P., and Faure G., Radiation from an equilibrium $CO_2$–$N_2$ plasma in the [250–850 nm] spectral region: II. Spectral modeling. Plasma Sources Science and Technology, 17, 3, pp.035013. (2008)

27. Mahmoudian A., Scales, W. A., Bernhardt P. A., Briczinski S. J, and McCarrick M. J., Investigation of ionospheric stimulated Brillouin scatter generated at pump frequencies near electron gyroharmonics. Radio Science, 48, 6, pp.685-697. (2013)

28. Unruh, W. G. "Experimental black-hole evaporation?. Physical Review Letters, 46,2,pp.1351. (1981)

29. Steinhauer, J. Observation of self-amplifying Hawking radiation in an analogue black-hole laser. Nature Physics, 10,11 pp.864, (2014)

30. Yilmaz, M. F., Tanrıseven, M. E., Obonyo, E., & Danisman, Y., Linear Discriminant Analysis As An Alternative Method To Investigate The Interaction Of A 1064 Nm Cw Laser Light With A Cold Inductively-Coupled Plasma. arXiv preprint arXiv:1802.05729, (2018)

31. Yilmaz M. F, Danisman Y., Larour J. and Arantchouk L., Linear discriminant analysis, based predator-prey analysis of hot electron effects on the X-pinch plasma produced K-shell Aluminum spectra. Scientific reports, 9, 1, pp.1-8. (2019)

32. Huba J. D, NRL plasma formulary. NAVAL RESEARCH LAB WASHINGTON DC PLASMA PHYSICS DIV. (2006)

33. Gheibi, A., Safari, H. and Innes D.E. Magnetoacoustic and Alfvénic black holes. The European Physical Journal C, 78,8, pp.662,( 2018).

34. Laurenza, M., Consolini, G., Storini, M. and Damiani, A.A Shannon entropy approach to the temporal evolution of SEP energy spectrum. Astrophysics and Space Sciences Transactions 8.1, pp.19-24. (2012)

35. Silbey, R. and Trommsdorff, H.P., Two-phonon relaxation in tunneling systems: An anomalous energy gap dependence. Chemical physics letters, 165.6, pp.540-544,(1990).

36. Leduc M., Julien D., and Juliette S., "Laser Cooling, Trapping, and Bose-Einstein Condensation of Atoms and Molecules." AIP Conference Proceedings, 1119, 1. (2009)

37. Gorbachev V. S and Kel'Ner S. R, Plasma condensation formation in a fluctuating strong magnetic field. Zh. Eksp. Teor. Fiz, 94, pp.89-99. (1988)

38. O'Shea E., Banerjee D. and Doyle J. G, Plasma condensation in coronal loops. Astronomy & Astrophysics, 475, 2, pp.L25-L28. (2007)

39. Maximov A. V, Oppitz R. M, Rozmus W., and Tikhonchuk V. T, Nonlinear stimulated Brillouin scattering in inhomogeneous plasmas. Physics of Plasmas, 7, 10, 4227-4237. (2000)

40. Salam, A. "Weak and electromagnetic interactions." Selected Papers Of Abdus Salam: (With Commentary), pp.244-25,(1994).

41. Hawking, S.W. Black hole explosions?. Nature 248,5443, pp.30-31. (1974)

42. Watt R. G., Cobble J., DuBois D. F., Fernandez J. C., Rose H. A., Drake R. P., and Bauer B.S., Dependence of stimulated Brillouin scattering on focusing optic F-number in long scale-length plasmas. Physics of Plasmas 3,3 pp.1091-1095 (1996).

43. Rahman H. U, Shukla P. K, and A. C, Das, Excitation of convection cells by ion-cyclotron waves. The Physics of Fluids, 24, 10, pp.1802-1805. (1981)

44. Stenzel R. L, Whistler waves with angular momentum in space and laboratory plasmas and their counterparts in free space. Advances in Physics: X, 1, 4, pp.687-710. (2016)

45. McKenzie J. F and Doyle T. B, Whistler oscillations and capillary-gravity generalized solitons. Quaestiones Mathematicae, 34, 3, pp.377-391. (2011)

46. Rabinovich V. S, and Sanchez I. M., Radiation from non-uniformly moving sources in the plasma. In Proceedings of the International Seminar Days on Diffraction, 2004, IEEE, pp.163-174. (2004)

47. Karlik, B., Yilmaz, M.F., Ozdemir, M., Yavuz, C.T. and Danisman, Y., A Hybrid Machine Learning Model to Study UV-Vis Spectra of Gold Nanospheres. Plasmonics, pp.1-9.( 2020.)

48. Sprangle P., Esarey E., and Ting A, Nonlinear theory of intense laser-plasma interactions. Physical Review Letters, 64, 17. (1990)